\documentclass[aps,pra,reprint,a4paper,superscriptaddress,floatfix,amsmath,amssymb,amsfonts,footinbib]{revtex4-1}

\usepackage[utf8]{inputenc}
\usepackage{graphicx}
\usepackage[normalem]{ulem}
\usepackage[usenames,dvipsnames]{color}
\usepackage[nointegrals]{wasysym}
\usepackage{commands}
\usepackage{upgreek}

\usepackage{amsmath}
\usepackage{placeins}
\usepackage[pdftex]{hyperref}
\hypersetup{pdfauthor={}, bookmarksnumbered=true, pdftitle={}, colorlinks, citecolor=blue, linkcolor=blue,urlcolor=blue}

\cleanversion
\renewcommand{\new}[1]{#1}

\begin{document}

\title{Optimization of ohmic contacts to n-type GaAs nanowires}

\author{L. H\"uttenhofer}\altaffiliation[Current address: ]{Center for NanoScience \& Fakult\"at f\"ur Physik, LMU-Munich, 80539 M\"unchen, Germany}\affiliationPDI
\author{D. Xydias}\affiliationPDI
\author{R.\,B. Lewis}\affiliationPDI
\author{S. Rauwerdink}\affiliationPDI
\author{A. Tahraoui}\affiliationPDI
\author{H. K\"upers}\affiliationPDI
\author{L. Geelhaar}\affiliationPDI
\author{O. Marquardt}\altaffiliation[Current address: ]{Weierstra{\ss}-Institut f\"ur angewandte Analysis und Stochastik, Mohrenstr.\ 39, 10117 Berlin, Germany}\affiliationPDI
\author{S. Ludwig}\email[Electronic address: ]{ludwig@pdi-berlin.de}\affiliationPDI

\date{\today}

\begin{abstract}
III-V nanowires are comprehensively studied because of their suitability for optoelectronic quantum technology applications. However, their small dimensions and the spatial separation of carriers from the wire surface render electrical contacting difficult. Systematically studying ohmic contact formation by diffusion to $n$-doped GaAs nanowires, we provide a set of optimal annealing parameters for Pd/Ge/Au ohmic contacts. We reproducibly achieve low specific contact resistances of $\sim2\times10^{-7}\,\Omega\text{cm}^2$ at room temperature becoming an order of magnitude higher at $T\simeq4.2\,$K. We provide a phenomenological model to describe contact resistances as a function of diffusion parameters. Implementing a transfer-matrix method, we numerically study the influence of the Schottky barrier on the contact resistance. Our results indicate that contact resistances can be predicted using various barrier shapes but further insights into structural properties would require a full microscopic understanding of the complex diffusion processes.  
\end{abstract}

\maketitle

\section{Introduction}
GaAs nanowires $n$-doped with silicon ($n$-GaAs) are promising building blocks for optoelectronic devices such as light emitting diodes (LEDs) and photovoltaic cells or for future quantum technology applications \cite{Yan2009,Czaban2009,Tomioka2010,Gutsche2011,Chuang2011,Borgstroem2011,Mariani2011,Krogstrup2013,Song2014,Miao2014,Orru2014,Dasgupta2014,Dimakis2014,Koblmueller2017}. 
For applications requiring charge transport or the integration into device circuitry, nanowires have to be electrically contacted \cite{Gutsche2011,Krogstrup2013,Song2014,Miao2014,Orru2014}. 
To achieve optimal device performance, \scrap{contact resistances need to be small and, within the range of interest, current independent, i.e.\ ohmic} \new{their contacts should have a small and current independent resistance, resembling ohmic contacts}. This imposes a challenge, as the band bending at metal-to-semiconductor contacts leads to Schottky barriers \cite{Schottky1939} reducing the carrier transmission and causing non-linear current-voltage ($I$-$V$) characteristics. A Schottky barrier forms by charge transfer between the semiconductor and the metal surface \new{driven by the equalization of} \scrap{equalizing} the chemical potentials of the materials \new{at the interface. This process results} \scrap{and resulting} in a charge depletion zone in the semiconductor \cite{Tung2014,Sze2007}. \new{The width of the depletion zone defines the Schottky barrier width $W_\text{SB}$. It depends on the concentration of free charge carriers in the semiconductor, governed by the doping concentration $n_\text D$, such that $W_\text{SB}\propto\sqrt{\phi_0/n_\text D}$. It can be tuned by doping, as increasing $n_\text D$ decreases $W_\text{SB}$ which enhances the transmission through the barrier by quantum tunneling.} The barrier height $\phi_0$ is a material-dependent constant which is related to the difference between the work function of the metal and the electron affinity of the semiconductor but, practically, strongly depends on defect states at the interface. \scrap{In contrast to its height, the barrier width $W_\text{SB}\propto\sqrt{\phi_0/n_\text D}$ depends on the doping concentration $n_\text D$ and, hence, can be tuned. Increasing $n_\text D$ decreases $W_\text{SB}$ and enhances the transmission through the barrier by quantum tunneling.} \scrap{Unfortunately,} Many applications require weak doping of semiconductors which yields wide Schottky barriers. \new{Unfortunately, such a wide barrier corresponds to} \scrap{with} a large specific contact resistance of typically $\sim 0.1\,\Omega\text{cm}^2$ or higher, even at room temperature. However, the specific contact resistance can be dramatically reduced and $I$-$V$ characteristics resembling ohmic behavior can be obtained by enhancing the doping locally, \scrap{where needed} \new{near the interface}. \scrap{Another} \new{An alternative} way to reduce the absolute contact resistance would be to increase the lateral extension of the contact region. This possibility is, however, limited if small nanostructures such as thin and short nanowires \scrap{are} \new{have to be} contacted. 

Low resistance ohmic contacts to planar $n$-GaAs wafers are frequently achieved by local diffusion of germanium (Ge) atoms into the semiconductor using rapid thermal annealing (RTA) \cite{Kim1997}. The detailed kinetics of such a diffusion process is material dependent, complex, and --in most practical cases-- not well understood. Consequently, optimizing contacts is often based on a trial-and-error approach. Improved contacts to $n$-GaAs had been achieved by adding a thin layer of palladium (Pd) placed between the GaAs surface and the Ge layer, where the Pd acts like a catalyst promoting the aspired diffusion of germanium into GaAs \cite{Marshall1987}. The specific diffusion process was described phenomenologically by a regrowth model \cite{Kim1997}. In a nutshell, the Pd enables out-diffusion of Ga from the GaAs wafer before it quickly diffuses through the Ge layer away from the GaAs surface \cite{Tahini2015}. The Ge atoms then diffuse into the gallium vacancies below the GaAs wafer surface.

Yielding low-resistive ohmic contacts to \new{GaAs nanowires \cite{Gutsche2011,Orru2014}} is generally more difficult than to bulk \new{GaAs} crystals because small wires allow only for small contact areas.\scrap{ while} \new{In addition,} their quasi one-dimensional geometry \new{with a high surface-to-volume ratio} alters the diffusion dynamics \new{in nanowires compared to bulk material}. A more aggressive diffusion (\eg\ at higher temperature), which in the case of bulk crystals often yields success, likely compromises the properties and functionality of the nanowire itself by excessive in-diffusion of Ge \new{(away from the contacts)} or \new{even} deterioration of the wire between the contacts (e.g.\ by evaporation of arsenic). Consequently, creating ohmic contacts to nanowires requires a particularly high level of control of the diffusion dynamics. The most successful approach to create ohmic contacts to $n$-GaAs nanowires so far is based on thermal annealing of layers of Pd, Ge and a capping layer of gold (Pd/Ge/Au) locally evaporated onto the wire \cite{Gutsche2011}. In the present work we performed an optimization of similar Pd/Ge/Au contacts by systematically exploring the annealing parameters: temperature and duration. Our wires have an undoped GaAs core covered by an $n$-doped GaAs shell. We achieved record low resistive ohmic contacts and confirm the enhanced quantum tunneling through the Schottky barriers with reduced width by transport measurements at liquid-helium temperatures.

Our systematic experimental results allow for a quantitative comparison with a phenomenological model, which describes the transmission through Schottky barriers as a function of diffusion parameters. Owing to a lack of knowledge regarding the detailed diffusion kinetics, we radically simplify our model to a single-stage process described by Fick's laws. As a consequence, we cannot predict the \new{realistic} shape of the Schottky barrier. As a way out, we model various barrier shapes. We find that the measured contact resistances can be predicted independent of the details of the Schottky barrier while it is impossible to extract microscopic details such as the Schottky barrier shape or diffusion parameters from our transport data. This would require a full understanding of the complex (multistep) microscopic diffusion dynamics.

\section{Experimental Details}

\begin{figure}[ht]
\includegraphics[width=1\linewidth]{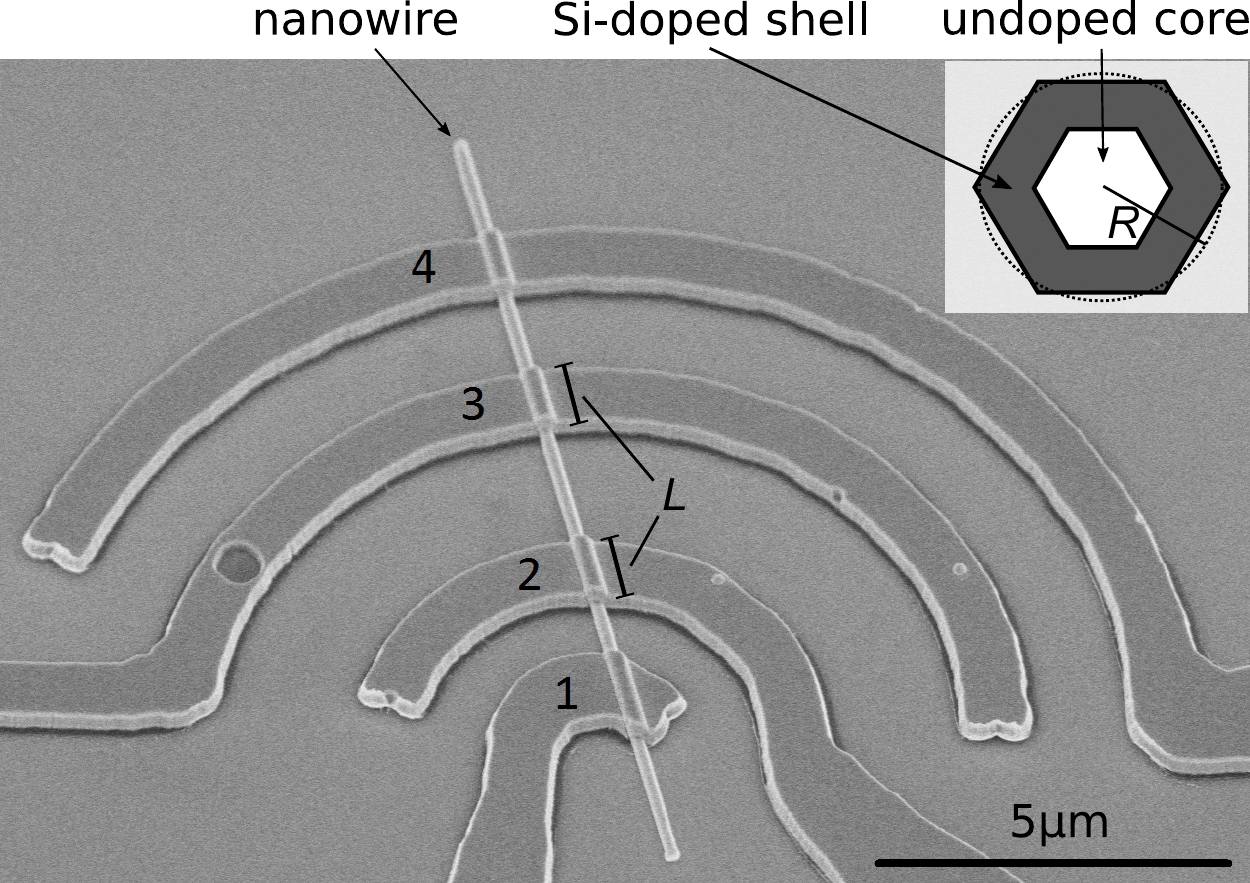}
\caption{SEM image of a nanowire including four contacts before annealing; each contact covers the nanowire over a length of $L\simeq1\,\mu$m. The leads form half circles to facilitate the contacting of statistically oriented nanowires by orienting the substrate relative to a mask for optical lithography. The inset sketches the cross section of a wire including GaAs core and $n$-GaAs shell. Approximating the hexagonal cross section by a cylinder, we can define the nanowire radius $R\simeq90$\,nm.
}
\label{fig:nanostructure}
\end{figure}

We have fabricated GaAs wires by Ga-assisted molecular beam epitaxy on a silicon (111) substrate covered by native oxide. After growing the $\simeq$\,$12\,\mu$m long and $\simeq$\,$95$\,nm wide undoped GaAs cores at $630\,^\circ$C, we added a $\simeq$\,$40\,$nm thick silicon-doped shell by lateral growth at $430\,^\circ$C, see Refs.\ \cite{Kuepers2017,Dimakis2012} for details. The nominal doping concentration in the shell is $c_\text{Si}=9\times10^{18}\,\text{cm}^{-3}$, determined for similar samples. For our transport experiments, we transferred the wires after growth (solute in isopropanol) to an $n^+$-doped silicon wafer, which is used as an electrical \new{field effect} back gate at the same time. To isolate the wires from the back gate, the wafer is covered with 50\,nm thermal oxide. For the fabrication of \new{the} Pd/Ge/Au \scrap{leads} \new{contacts to the nanowire,} we used optical lithography.\scrap{ to pattern the nanowires on the surface.} \new{After development} the samples were \scrap{then} cleaned in a HCl:H$_2$O solution at the ratio 1:200 for 20\,s, afterwards rinsed in DI water, blown dry with N$_2$ and loaded promptly into an electron-beam evaporation chamber. \scrap{In-situ, Prior the metal deposition,} \new{Within the load lock of the evaporation chamber,} we applied three minutes of Argon-ion sputtering at 36\,eV.  \scrap{on the developed parts of the nanowire to remove the native oxide on the surface} \new{This way the native oxide is removed from the sample, where not covered by resist, including the nanowire surface at the contact positions}. Finally, the metal layers, namely 50\,nm of Pd, 170\,nm of Ge, and 80\,nm of Au, were deposited followed by a metal lift-off process. In \onefig{fig:nanostructure},
we present a scanning electron microscope (SEM) image of a typical wire with four contacts.

Aiming at quantum applications, we are interested in the low-power properties of the nanowires. Consequently, we characterized the nanowires by measuring $I$-$V$ curves for $V<80\,$mV employing both, two-terminal (2TM) and four-terminal (4TM) transport measurements. \scrap{Here, we discuss} \new{In our} 2TM measurements \scrap{with} \new{we applied a constant voltage} $V$ \scrap{applied} between the inner contacts, 2 and 3, \scrap{as shown in} \new{see} \onefig{fig:nanostructure}, and \new{measured current} $I$ \scrap{measured} using a current voltage amplifier. \scrap{ as well as} \new{In our} 4TM measurements \scrap{with} \new{we applied} \new{a constant} $I$ \scrap{applied} between the outer contacts, 1 and 4, and  \new{measured} $V$ \scrap{measured} between the inner contacts. Assuming that the 4TM resistance $R_\text{4TM}\new{=R_\text{NW}}$ represents the nanowire resistance between the inner contacts and neglecting much smaller lead resistances, we can express the average of the two contact resistances (at contacts 2 and 3) as $R_\text c\simeq (R_2+R_3)/2\simeq 1/2 (R_\text{2TM}-R_\text{4TM})$. Evaporation of the contacts onto a hexagonal shaped nanowire suffers from a shadowing effect such that in our case only roughly 50\% of the nanowire circumference is actually in direct contact with the contact material. \scrap{From that} \new{Under this assumption}, we can estimate the specific contact resistance as $\rho_\text c\simeq\pi RLR_\text c\simeq0.27\,\mu\text m^2R_\text c$, where $R=90\,$nm and $L=1\,\mu$m are the nominal radius of the wire and width of our contacts, respectively.

\begin{figure}[ht]
\includegraphics[width=0.7\linewidth]{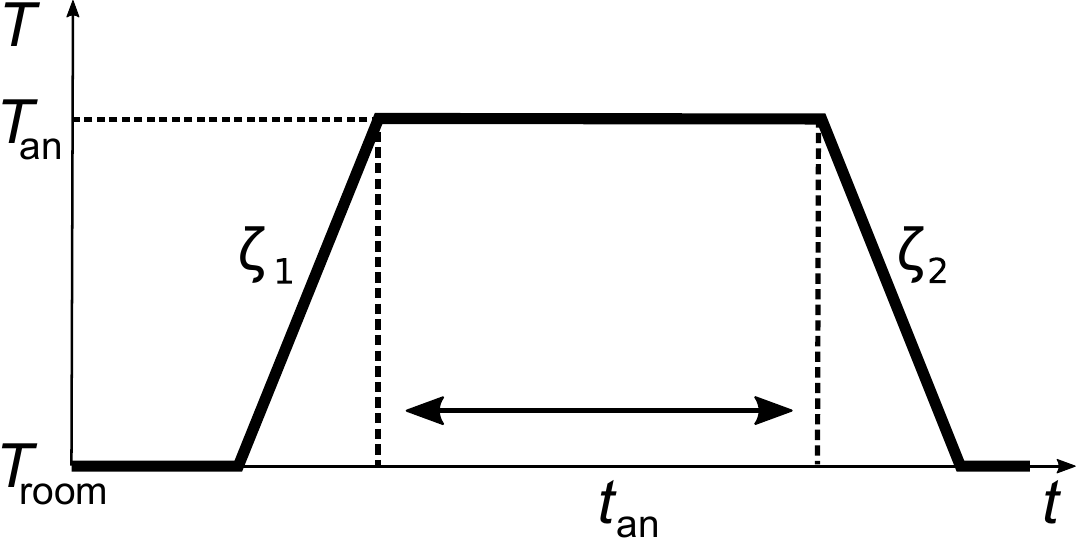}
\caption{Temperature versus time during annealing in a Dr. Eberl AO500 RTA-furnice. Temperature is ramped with the rate $\zeta_1=5\text{ K}/\text{s}$ and then kept constant at $T_\text{an}$ for the annealing duration $t_\text{an}$. Then the sample is cooled down to room temperature with the rate $\zeta_2\simeq-\zeta_1$.}
\label{fig:RTA}
\end{figure}

For comparison, we measured the 2TM and 4TM $I$-$V$ characteristics before and after diffusion of the contacts by RTA in a furnace with electrical heating and nitrogen flow cooling such that the heating and cooling rates where roughly equal and constant at $\pm5$\,K/s, as sketched in \onefig{fig:RTA}.
To systematically study the diffusion process, we varied the annealing time $t_\text{an}$ and temperature $T_\text{an}$ for various samples.

\section{Experimental Results}

\begin{figure}[ht]
\includegraphics[width=0.8\linewidth]{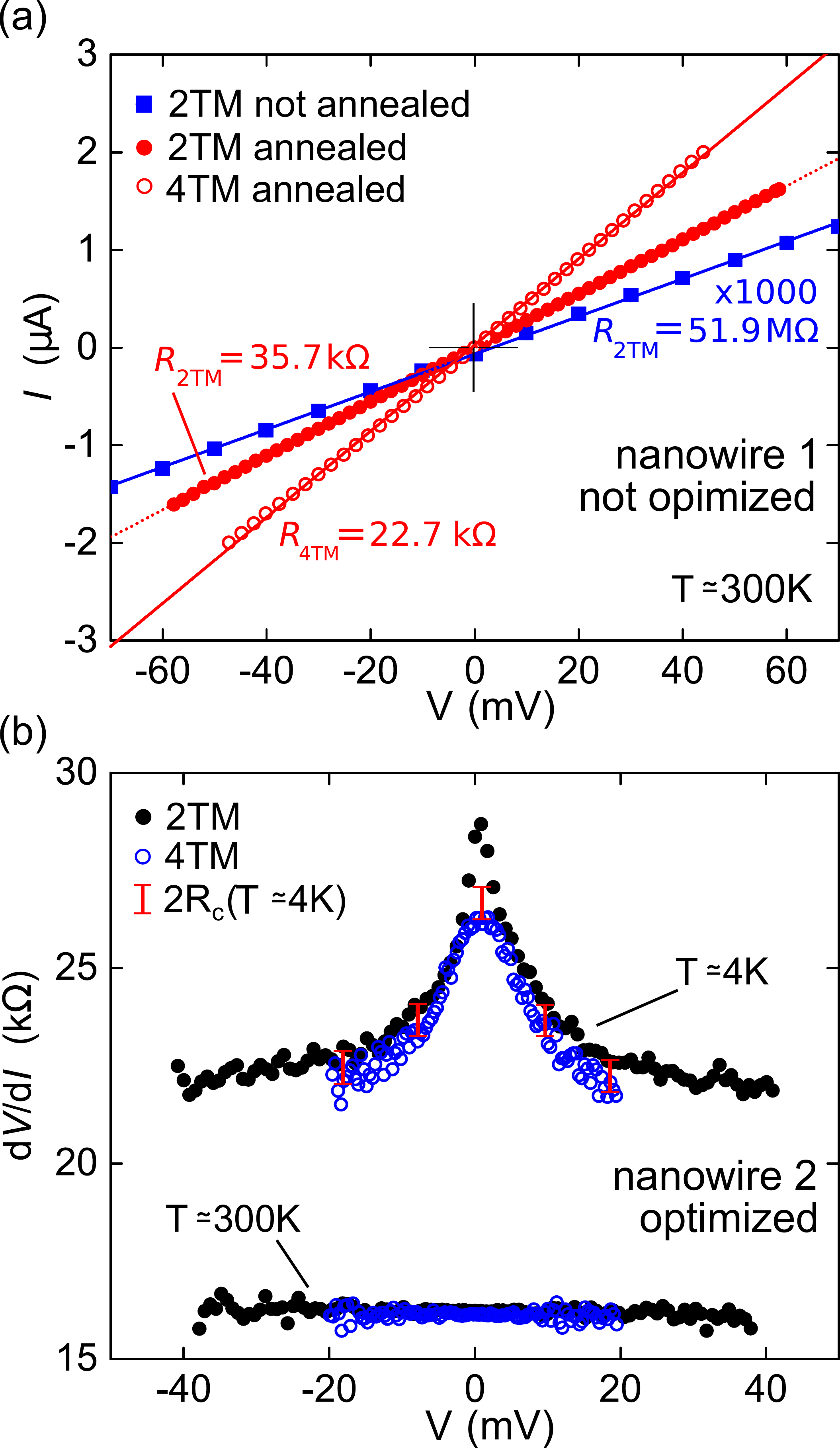}
\caption{a) Room temperature $I$-$V$ characteristics of a GaAs nanowire before and after RTA. While not yet optimized ($t_\text{an}=30$\,s, $T_\text{an}=270\,^\circ$C), RTA reduces the contact resistance by a factor of 4000 from $2R_\text C\simeq52\,\text M\Omega$ to about $13\,\text k\Omega$. 
b) Derivative of $I$-$V$ curves (differential resistance) after optimal RTA ($t_\text{an}=90$\,s, $T_\text{an}=270\,^\circ$C). Comparison of room temperature and cryogenic characteristics. The red scale bars indicate a contact resistance of $2R_\text C\simeq1\,\text k\Omega$ at 4.2\,K, while  $2R_\text C\simeq0.1\,\text k\Omega$ at room temperature.
}
\label{fig:I-V}
\end{figure}

In \fig{fig:I-V}a,
we show typical $I$-$V$ curves resulting from 2TM and 4TM measurements at room temperature. All curves origin from the same nanowire. Before RTA (\textit{not annealed}), $R_\text{2TM}\simeq52\,\text M\Omega$ is entirely dominated by the contacts. After not yet optimized annealing at $T_\text{an}=270\,^\circ$C for $t_\text{an}=30$\,s, the contact resistance drops by a factor of 4000 to $2R_\text C=R_\text{4TM}-R_\text{2TM}\simeq13\,\text k\Omega$, while the nanowire resistance stays unaffected at $R_\text{4TM}\simeq22.7\,\text k\Omega$ ($R_\text{4TM}$ before annealing is not shown). In order to determine the parameters for an optimal annealing procedure, we carried out a systematic investigation. Before discussing details, we present in \fig{fig:I-V}b the characteristics after optimal annealing. \scrap{including} \new{The figure includes} 2TM and 4TM measurements at both, cryogenic and room temperature. To reveal more details, here we plot the differential resistance, i.e.\ the derivative of $I$-$V$ curves, instead of current versus voltage. The constant value of $\text d V/\text dI$ at $T\simeq300\,$K corresponds to a linear $I$-$V$ characteristic and we find very low contact resistances of $2R_\text C\simeq100\,\Omega$. Upon cooling to $T\simeq4.2\,$K in a liquid helium bath, the wire resistance ($R_\text{4TM}$) as well as the contact resistances $2R_\text C=R_\text{4TM}-R_\text{2TM}\simeq\,1\text k\Omega$ increase \footnote{\new{This increase of contact resistances while lowering the temperature is an intrinsic feature of a Schottky barrier. It reflects the energy dependence of transmission through a barrier of finite height and has been previously discussed in terms of a transition from classical thermionic emission to quantum mechanical field emission \cite{Kim1997,Tung2014}. Our model reflects the temperature dependence by integrating over the carrier excitation spectrum in the leads, see Appendix for details.}}. Nevertheless, we have achieved contact resistances which are even at cryogenic temperatures an order of magnitude lower than our wire resistances and, importantly, much lower than the fundamental one-dimensional resistance quantum of $h/e^2\simeq25.8\,\text k\Omega$. Such low contact resistances would allow for ballistic quantum transport measurements.

The low-temperature measurements reveal a slightly non-ohmic characteristics at low energies: at 4.2\,K the nanowire resistance $R_\text{4TM}$ increases by 18\% as $V$ is decreased from 20\,mV to zero. We interpret this behavior as a signature of disorder in our wires, which causes hopping transport (caused by disorder-induced barriers) at cryogenic temperatures. This excludes the possibility of ballistic linear-response transport in these specific wires. Interestingly, we find \scrap{that $R_\text{2TM}$ increases much stronger than} \new{a particularly strong increase of $R_\text{2TM}$ compared to} $R_\text{4TM}$ for very small $V\lesssim1\,$mV, \scrap{such that the contact resistance doubles} \new{visible as a sharp maximum of the low temperature $R_\text{2TM}$, black dots in \fig{fig:I-V}b.} While this observation requires further investigation, it might be attributed to a disorder potential increasing the effective Schottky barrier width at low energies, i.e.\ for $eV<\kT$. We expect that disorder-free nanowires would allow for ohmic $I$-$V$ characteristics at cryogenic temperatures and very low energies \footnote{Disorder in our specific nanowires can be caused by polytypism or surface states, but it is likely dominated by the Coulomb potential of donor ions as the free carriers reside in the same material layer.}. 

\begin{figure}[]
\includegraphics[width=1\linewidth]{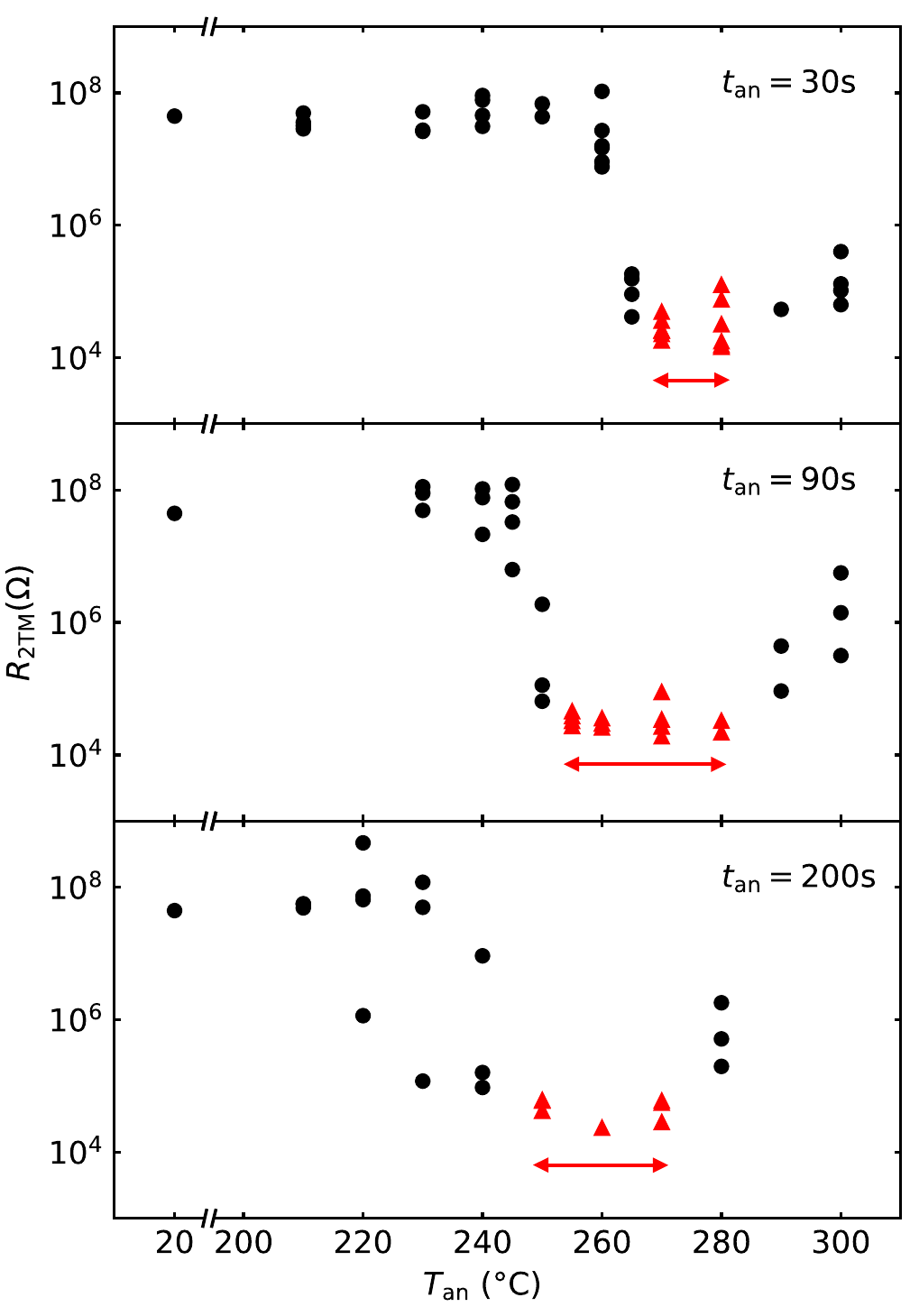}
\caption{$R_\text{2TM}$ as a function of annealing temperature $T_\text{an}$ for various annealing durations $t_\text{an}$. Optimal ranges for yielding low contact resistances are indicated by \new{red triangles} and double arrows. Individual data points correspond to individual nanowires with the exception that some nanowires have been measured at two different $T_\text{an}$ values if the first annealing process did not yet yield a reduced resistance. This has a negligible influence on the results because of the strong temperature dependence of the diffusion dynamics.}
\label{fig:RvsT}
\end{figure}

\begin{figure}[ht]
\includegraphics[width=0.8\linewidth]{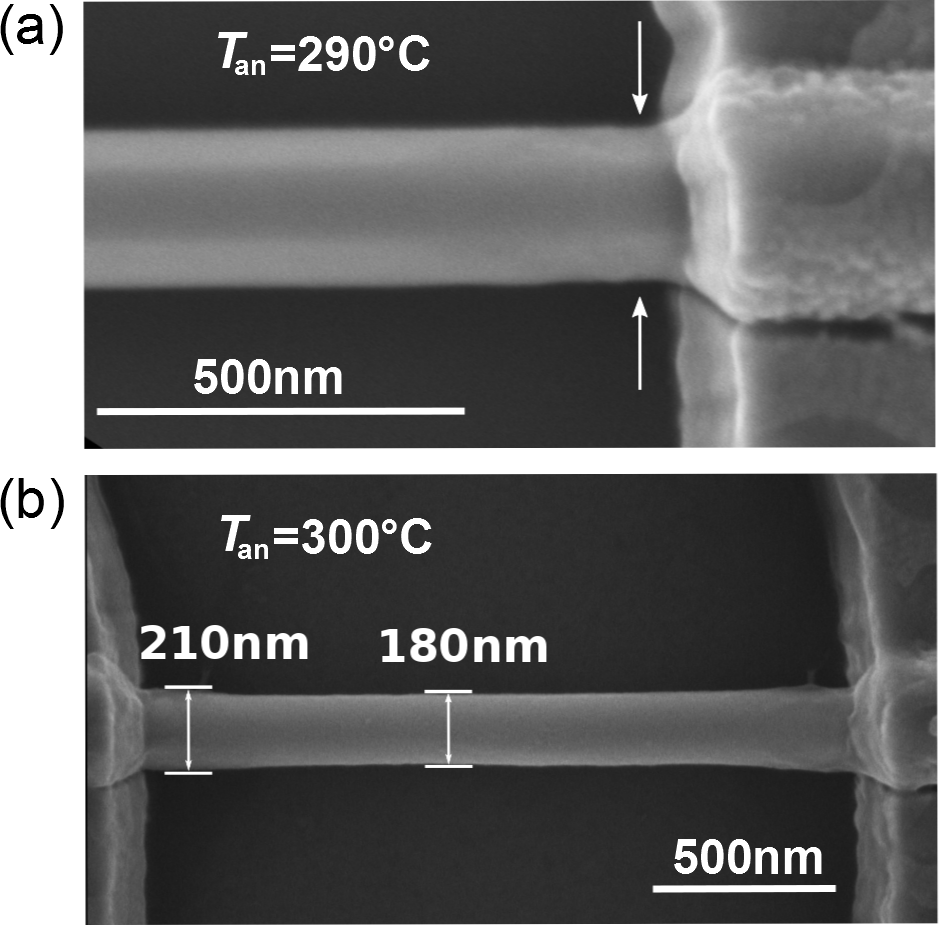}
\caption{SEM images of segments of two contacted nanowires after RTA with $t_\text{an}=90\,$s but $T_\text{an}$ exceeding the temperature range for \new{optimal} ohmic contact formation.  a) At $T_\text{an}=290^\circ$C the nanowire starts to thin by evaporation of arsenic in direct vicinity of the metal contacts, indicated by arrows. (b) As $T_\text{an}$ is further increased to 300$^\circ$C, \scrap{the measured diameters diminish towards the segment center. This indicates that the deterioration of the nanowire now extends over the whole segment and becomes stronger away from the contacts. Scale bars correspond to 500\,nm.} \new{the nanowire deterioration becomes stronger, in particular away from the contacts, where the measured diameter (indicated) decreases strongly.}}
\label{fig:deterioration}
\end{figure}

Searching for the optimal annealing procedure, we systematically varied the RTA temperature $T_\text{an}$ and duration $t_\text{an}$. In \fig{fig:RvsT}{},
we present the resulting $R_\text{2TM}(T_\text{an})$ for three different $t_\text{an}$, all measured at room temperature. The highest measured resistances of $R_\text{2TM}\sim50\,\text M\Omega$ correspond to contacts before annealing. The lowest measured resistances of $R_\text{2TM}\simeq25\,\text k\Omega$ are dominated by the nanowire resistance and correspond to contact resistances of $2R_\text C\sim 100\,\Omega$. We generally observe a rapid drop in contact resistance as $T_\text{an}$ is increased, where the drop happens at higher temperature for shorter annealing duration, in qualitative agreement with Fick's diffusion laws. Importantly, on the one hand, faster annealing at higher temperature yields more reproducible results within the relevant range of parameters. (The scattering of $R_\text{2TM}(T_\text{an})$ near the drop increases for longer $t_\text{an}$.) On the other hand, as $T_\text{an}$ is increased beyond $280\,^\circ$C, the \new{measured} resistance grows again. We believe that this resistance increase is related to evaporation of arsenic at high temperatures which can cause a structural deterioration of the nanowires \cite{Gutsche2011}. SEM images presented in \fig{fig:deterioration}{}
support our explanation \footnote{In our transport measurements as well as in energy dispersive X-ray spectroscopy (EDX) measurements (not shown) we find no indications of an enhanced germanium concentration in the nanowires away from the contact areas.}. 

As a result, the optimal range for RTA, which is indicated in \fig{fig:RvsT}{} by red double arrows, is limited by the initial resistance drop from low $T_\text{an}$ and the final resistance increase at high $T_\text{an}$. It is immediately evident, that our intermediate $t_\text{an}=90\,$s allows for the largest temperature window which promises lowest possible contact resistances.

\section{Model}

In the following, we develop a phenomenological model to describe the (room temperature) contact resistance as a function of annealing parameters $R_\text C(T_\text{an},t_\text{an})$. \scrap{Such a model could} \new{The purpose of the model is to} provide additional understanding and help us to further improve the contacts to III-V nanowires. The model consists of two steps: first, the identification of the donor distribution after annealing which determines the width of the Schottky barrier and, second, the computation of the resistance of the Schottky barrier, i.e.\ the contact resistance.

\begin{figure}[ht]
\begin{center}
\includegraphics[width=1\linewidth]{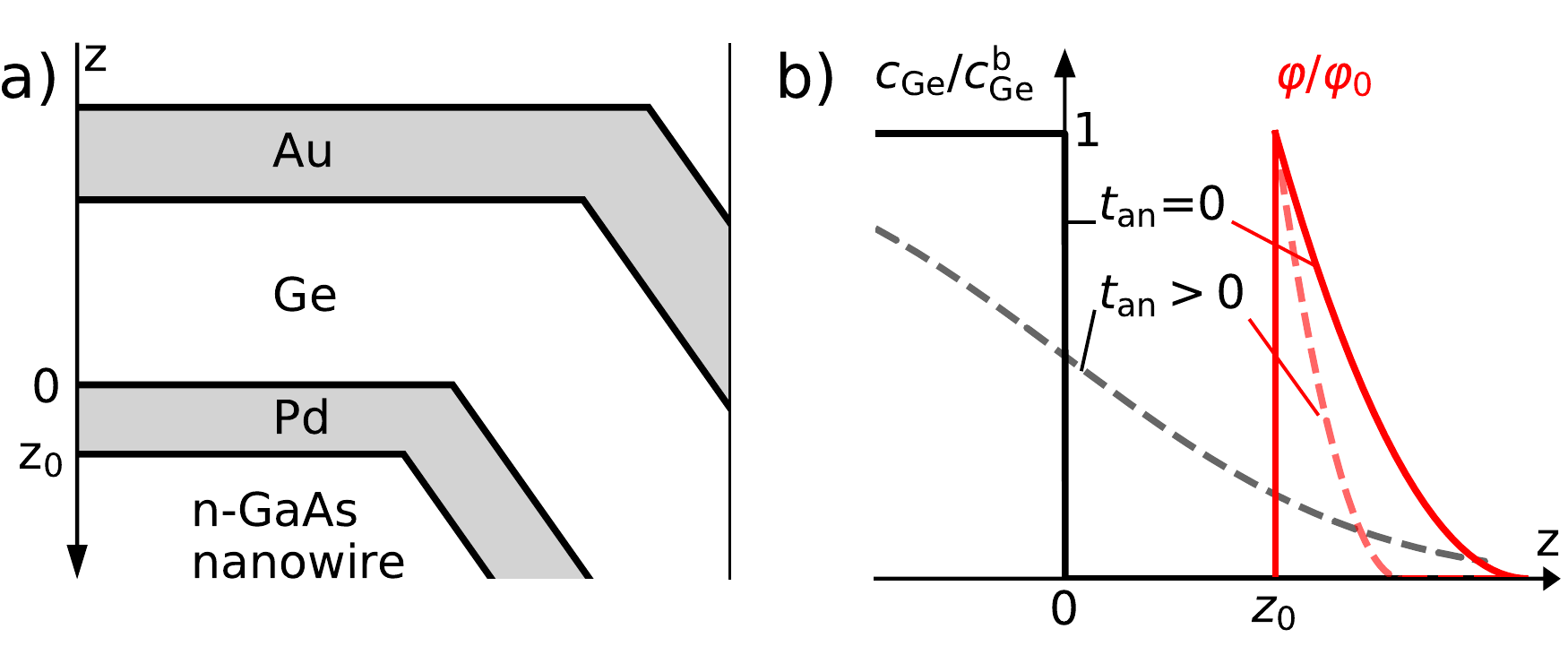}
\end{center}
\caption{\new{a) Material stacking at contacts to the GaAs nanowire before annealing. The sketch shows two facets of the hexagonal nanowire, compare inset of Fig.\ \ref{fig:nanostructure}.
b) Illustration of the simplified one-dimensional diffusion model: Before annealing, i.e.\ for $t_\mathrm{an}=0$, the germanium concentration is approximated in terms of two half-spaces, where $c_\mathrm{Ge}(z<0)=c_\mathrm{Ge}^{\mathrm{b}}$, assuming pure germanium, while $c_\mathrm{Ge}(z\ge0)=0$ (solid black line). After annealing for $t_\mathrm{an}>0$ the germanium concentration is altered by diffusion (dashed black line). Germanium that diffused into the nanowire reduces the width of the Schottky barrier at the metal-semiconductor interface. The Schottky barrier is sketched by solid versus dashed red lines before and after annealing.}}
\label{fig:diff_crs}
\end{figure}

The diffusion dynamics in our layered system containing several materials is very complicated and goes beyond the scope of the present paper. Here, we radically simplify the complex dynamics by assuming a one-dimensional diffusion process of Ge into the semiconductor described by Fick's laws. \new{Our simplified one-dimensional diffusion model is illustrated in \fig{fig:diff_crs}{}.}
Using separate half spaces as \scrap{boundary} \new{initial} condition \new{before annealing}, we assume the \new{initial ($t_\mathrm{an}=0$)} concentration \new{distribution} $c_\text{Ge}(z>0)=0$ and $c_\text{Ge}(z\le0)=c^\text{b}_\text{Ge}$ \scrap{at $t=0$} with $c^\text{b}_\text{Ge}$ being the bulk atomic concentration \new{within the layer} of pure Ge\new{, see \fig{fig:diff_crs}{a}}. \scrap{Within this approximation, we can estimate} \new{Applying Fick's laws we model} the Ge concentration near the nanowire surface as a function of diffusion depth $z$, annealing time $t_\mathrm{an}$, and annealing temperature $T_\mathrm{an}$ as \cite{Gottstein2007}
\begin{equation}\label{eq:c}
c_\text{Ge}(z)=\frac{c^\text{b}_\text{Ge}}{2}\left[1+\text{erf}\left(\frac{-z}{2\cdot\sqrt{t_\text{an}D_0\exp\left(\frac{-E_\text{A}}{k_\text{B}T_\text{an}}\right)}}\right)\right]\,,
\end{equation}
where $D_0$ and $E_\text{A}$ denote the diffusion coefficient and hopping activation energy, respectively\scrap{, according to Fick's laws}. The nanowire doping concentration predicted by \eq{eq:c} is then 
\begin{equation}\label{eq:nd}
n_\text{D}(z)=c_\text{Si}+c_\text{Ge}(z)
\end{equation}
assuming that every incorporated Ge atom and every Si atom act equally as one donor atom \footnote{\new{Note, that we neglect compensation effects which can arise if e.g.\ a fraction of silicon dopants act as acceptor instead of donor. It has been shown, that such compensation effects are minimal in our core-shell wires \cite{Dimakis2012}.}}.

As a consequence of the above simplification for the diffusion dynamics, we cannot predict the realistic microscopic donor distribution in detail. Consequently, the shape of the Schottky barrier is also not precisely known. As hinted above, this \scrap{is a very general problem} \new{limitation is very common} and, in particular, applies to nanostructures with Schottky barriers fabricated by local diffusion. In order to estimate the consequences of an unknown Schottky barrier shape and to determine the relevance of a specific shape for the physical predictions of a model, we calculate the transmission for various barrier shapes.

In our first calculation, we follow the standard approach by assuming a uniform $n_\text D$ using $c_\text{Ge}(z=z_0)$ with $z_0=50\,$nm as a reference depth in \eq{eq:c}. Furthermore, we neglect screening effects such that the Schottky barrier takes its text book shape
\begin{equation}\label{eq:schott}
\phi(z)=\phi_0\left(1-\frac{z}{W_\text{SB}}\right)^2\\
\end{equation}
for $0\le z\le W_\text{SB}$ and $\phi(z)=0$ elsewhere
\new{and with the barrier width determined by}
\begin{equation}
W_\text{SB}=\sqrt{\frac{2\varepsilon_0\varepsilon_\text{GaAs}}{en_\text{D}}\phi_0}\,.\nonumber
\end{equation}
Here, $\varepsilon_0$ and $\varepsilon_\text{GaAs}$ are the dielectric constant of vacuum and GaAs, respectively, while $e$ is the elementary charge. The barrier height $\phi_0$ is, in first approximation, a material constant independent of $n_\text{D}$, hence independent of the detailed barrier shape.

To capture thermally activated hopping over the barrier as well as quantum tunneling, the transmission through a Schottky barrier, \scrap{which is often assumed to have the shape} described by \eq{eq:schott}, is traditionally approximated in terms of three separate contributions: thermionic emission, field emission, and thermionic field emission \cite{Sah1991,Kim1997,Sze2007}. The accuracy of this semiclassical approach is hard to predict away from the extreme limits (deep classical or quantum limit). For these limits, the barrier shape does not matter. For the case of a (truncated) parabolic barrier, the exact transmission can be calculated analytically, which was suggested as a possible route to approximate the transmission through realistic Schottky barriers \cite{Hansen2004}. Instead, we calculate numerically the full transmission for arbitrary barrier shapes. Our analysis below shows that physical predictions require knowledge of the realistic barrier shape. 

\begin{figure*}[t]
\begin{center}
\includegraphics[width=.85\textwidth]{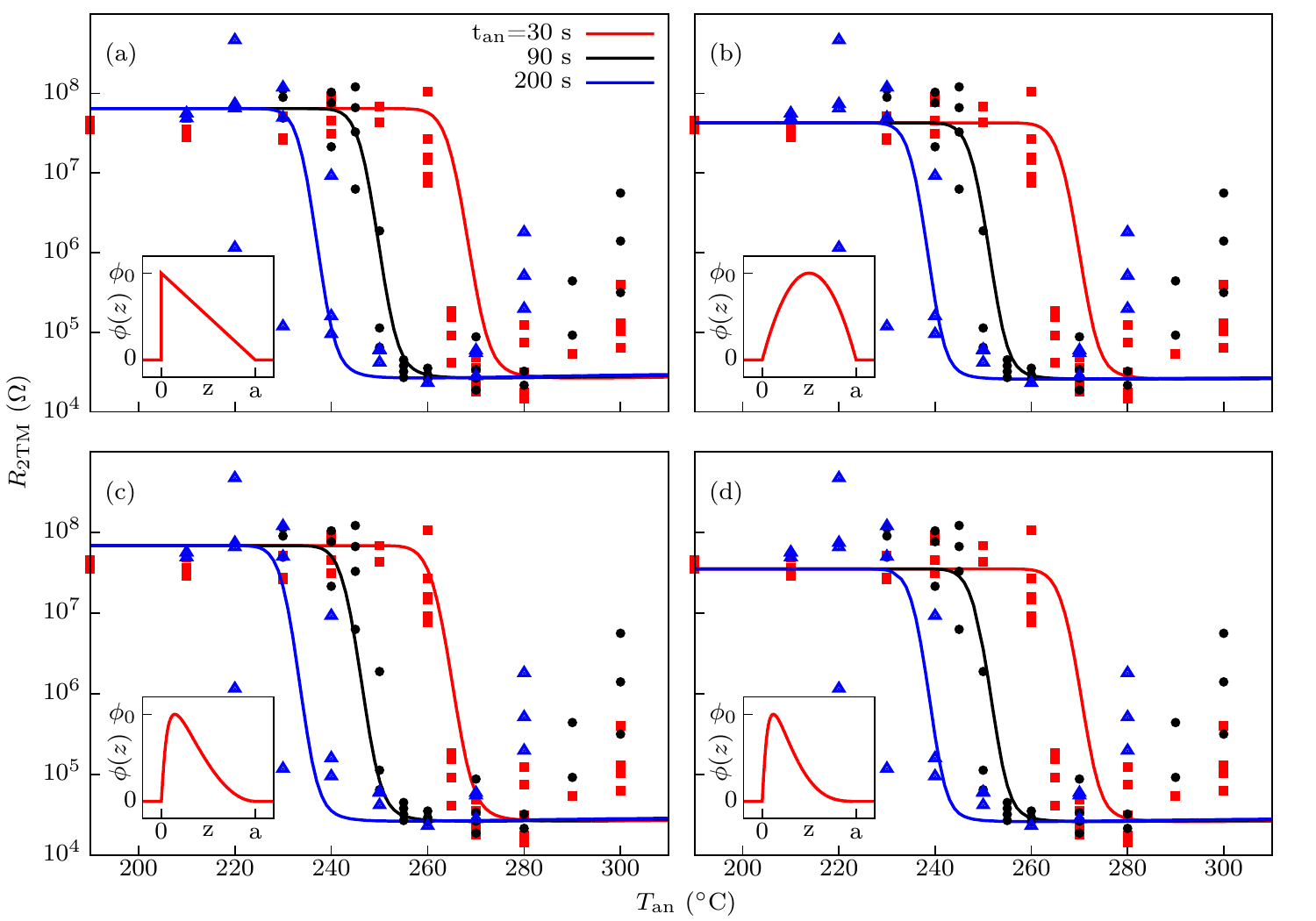}
\end{center}
\caption{$R_\text{2TM}(T_\mathrm{an})$ measured at $300$\,K for annealing times $t_\mathrm{an}=30$\,s (red), 90\,s (black), and 200\,s (blue). The data are identical to those in Fig. \ref{fig:RvsT} and are plotted as symbols. Model curves are plotted as lines. Potential shapes are shown in the insets. They are a) triangular
$\phi(z)=\phi_0\cdot(1-z/a)$,
b) parabolic $\phi(z)=\phi_0\cdot(1-4(z/a-0.5)^2)$,
c) half a parabola with softened maximum $\phi(z)=\phi_0\cdot(\arctan(15z/a)\cdot(1-z/a)^2)$, and
d) half a cubic parabola with softened maximum $\phi(z)=\phi_0\cdot(\arctan(15z/a)\cdot(1-z/a)^3)$.
All potentials are zero for $z<0$ and $z>a$.
}
\label{fig:rtot}
\end{figure*}

Implementing a transfer-matrix method (TMM) \cite{Tsu1973}, we have evaluated the \scrap{influence of} \new{contact resitance $R_\text C$ for} four different shapes of the Schottky barrier, as illustrated in the insets of \fig{fig:rtot}{}.
A detailed description of the numerical model is given in the Appendix. Our first example is a simple triangular potential, $\phi(z)=\phi_0\cdot(1-z/a)~\forall~0 \le z \le a$. The second example is a parabolic barrier with maximum at a/2, $\phi(z)=\phi_0\cdot(1-4(z/a-0.5)^2)~\forall~0 \le z \le a$. As the parabolic barrier allows for an analytical calculation \cite{Hansen2004}, we take it as a reference to verify our numerical calculations, as discussed in the Appendix. Next, we have assumed half of a parabola with a maximum softened by an arctangent function, $\phi(z)=\phi_0\cdot(\arctan(15z/a)\cdot(1-z/a)^2)~\forall~0 \le z \le a$. This potential shape is very close to (albeit more realistic than) the Schottky barrier for a homogeneous carrier distribution described by \eq{eq:schott}. Finally, to account for the expected decrease of the donor concentration away from the nanowire surface as predicted by \eq{eq:c}, we have employed a steeper potential resembling half a cubic parabola with a softened maximum, $\phi(z)=\phi_0\cdot(\arctan(15z/a)\cdot(1-z/a)^3)~\forall~0 \le z \le a$. We expect the last one to describe the most realistic barrier shape.

The simulation parameters, i.e.\ height of the Schottky barrier $\phi_0$, activation energy $E_\text a$, and the diffusion constant $D_0$ are not well known. Together, they define the barrier width $a$, assuming $a=W_\text{SB}$ via \eqs{eq:c}--(\ref{eq:schott}). We determine these parameters from a least-squares fit of the simulations to measured \new{2TM} resistances\new{, $R_\text{2TM}=2R_\text C+R_\text{NW}$. In \fig{fig:rtot}{} we present the calculated curves $R_\text{2TM}(T_\text{an})$ in comparison to our measured data points already seen in \fig{fig:RvsT}{}. For our calculations} we used $n_\text D(z_0)$ with $z_0=50\,$nm, which is the thickness of the palladium layer, as germanium has to pass the palladium before diffusing into the GaAs nanowire. This choice reflects the simplification of a complex diffusion process to a one-dimensional model described by Fick's laws. A different $z_0$ will change the absolute values of $E_\mathrm{a}$ and $D_\mathrm{0}$, but does not affect their relative changes and our main conclusions. The parameters \scrap{fitted for} \new{thus obtained by fitting} the four potential shapes are listed in Tab.~\ref{tab:fit}.
\begin{table}[ht]
	\begin{tabular}{c|ccc}
		$\phi(z)/\phi_0$ & $\phi_0$ (meV) & $E_\mathrm{a}$ (eV) & $D_\mathrm{0}$ (m$^2$/s)\\
		\hline
		$1-z/a$ & 679 & 1.440 & 8.46$\cdot$10$^{-5}$\\
		$1-4(z/a-0.5)^2$ & 583 & 1.442 & 7.92$\cdot$10$^{-5}$\\
		$\arctan(15z/a)\cdot(1-z/a)^2$ & 781 & 1.408 & 5.10$\cdot$10$^{-5}$\\
		$\arctan(15z/a)\cdot(1-z/a)^3$ & 816 & 1.440 & 7.42$\cdot$10$^{-5}$
	\end{tabular}
\caption{Fit parameters obtained from a least-squares fit for the considered potential shapes and used for the model curves presented in Fig.\ \ref{fig:rtot}.}
\label{tab:fit}
\end{table}
\scrap{$E_\mathrm{a}$ and $D_\mathrm{0}$ are almost identical for the first three barrier shapes, but differ for the cubic barrier shape. The fitted barrier heights $\phi_0$ differ significantly for the various barrier shapes.
The respective resistance curves $R(T_\mathrm{an})$ are shown in \fig{fig:rtot}{}.
The model curves reproduce the measured data equally well, independent of the barrier shape.}
\new{While $E_\mathrm{a}$ remains quite similar for all potential shapes under consideration, $\phi_\mathrm{0}$ and $D_\mathrm{0}$ exhibit stronger variations. As demonstrated in \fig{fig:rtot}{} within the measurement accuracy our model captures the measured two-terminal resistance as a function of annealing temperature and time correctly, independent of the details of the Schottky barrier shapes. As a result, the specific shape of the Schottky barrier clearly cannot be concluded from our simulations.

The reduction of $R_\text{2TM}$ observed with increased $T_\mathrm{an}$ in Figs.\ \ref{fig:RvsT} and \ref{fig:rtot} reflects the dependence of the barrier width W$_\mathrm{SB}$ on $T_\mathrm{an}$ and t$_\mathrm{an}$ [also see discussion of \fig{fig:RvsT}{} above].}
\scrap{The fact that the horizontal shifts of the resistance drops in the $R_\text{2TM}(T_\text{an})$ curves for different $t_\text{an}$ are captured well by our model,}
\new{Our model curves capture the resitance drop by more than three orders of magnitude and the dependence of this drop on both, the annealing temperature $T_\text{an}$ and duration $t_\text{an}$ correctly. This} supports our procedure of simplifying a complex diffusion process by Fick's laws. A detailed physical understanding beyond our analysis, a prediction of the contact resistance, or a prediction of the perfect annealing parameters would require the knowledge of the exact shape of the Schottky barrier. However, if such an insight can be achieved, the TMM is a reliable and computationally efficient method to provide further \new{quantitative} information\new{, e.g.\ on the barrier width and height, which then could lead to a further optimization of the electrical contacts to nanowires}.

\begin{figure}[ht]
\begin{center}
\includegraphics[width=1\linewidth]{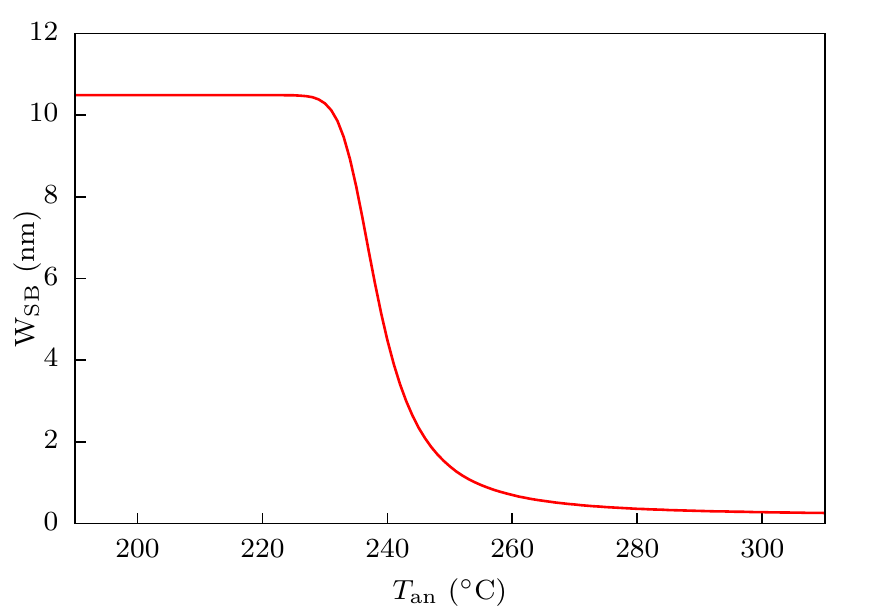}
\end{center}
\caption{$W_\mathrm{SB}$ 
of the barrier described by half a cubic parabola as a function of the annealing temperature $T_\mathrm{an}$ for $D_\mathrm{0}$=6.004$\times$10$^{-5}$m$^2$/s, $E_\mathrm{a}$=1.424~eV and $t_\text{an}=200\,$s.}
\label{fig:aVsT}
\end{figure}

For completeness, we present the Schottky barrier width dependence on annealing temperature $W_\text{SB}(T_\text{an})$ in \fig{fig:aVsT}{}
calculated with \eqs{eq:c}--(\ref{eq:schott}).

\section{Summary}
We studied the effect of annealing parameters on the formation of ohmic contacts to GaAs nanowires and thereby minimized the contact resistance. Our procedure allows to systematically and reliably optimize ohmic contacts by diffusion and can be applied to arbitrary nanostructures. We developed a phenomenological model describing the formation of a Schottky barrier by donor diffusion and determining the transmission through the barrier using the transfer-matrix method. The latter accurately describes our data independent of the shape of the Schottky barrier. In order to precisely predict physical parameters it would be necessary to know the shape of the barrier, which --in turn-- would require a detailed understanding of the complex diffusion processes on a microscopic level.

\section{Acknowledgments}
We would like to acknowledge technical help from Walid Anders, Anne-Kathrin Bluhm, Uwe Jahn, Angela Riedel and Werner Seidel.

\section{Appendix}

\subsection*{Numerical transfer matrix method}
A particle with energy $E$ is transmitted from left to right through a barrier with transmission $\Theta$. We begin with considering the one-dimensional problem and thereby omit the time dependence, which is not relevant for our problem. The wave function of the particle impinging the barrier from the left is then $\Psi_1(z)=A_1\text e^{ik_1 z}$, which is partially reflected to the left with amplitude $\Psi_\mathrm{R}(z)=A_\mathrm{R}\text e^{-ik_1 z}$ and transmitted to the right side of the barrier with $\Psi_\mathrm{T}(z)=A_\mathrm{T}\text e^{ik_2 z}$. The relationship between $E$ and wave numbers $k_{1,2}$ can be found by substituting these wave functions into the time independent Schr\"odinger equation
\begin{equation*}
\frac{\hbar^2}{2m}\frac{\partial^2\Psi(z)}{\partial z^2}=\Phi(z)\Psi(z)\,.
\end{equation*}
With $\Phi_\mathrm{l}$ and $\Phi_\mathrm{r}$ denoting the values of the potential $\Phi(z)$ at the left and right side of the barrier we find
\begin{equation*}
E_\mathrm{l}=E-\Phi_\mathrm{l}=\frac{\hbar^2k_1^2}{2m}\quad\text{and}\quad
E_\mathrm{r}=E-\Phi_\mathrm{r}=\frac{\hbar^2k_2^2}{2m}\,,
\end{equation*}
where $m$ is the effective mass which we assume to be constant, resembling the parabolic band approximation. To determine the wave function for a finite barrier $\Phi(z)$ of arbitrary shape between $z=0$ and $z=a$, we predefine the value of the wave function at $z=a$ to be unity and calculate its corresponding derivative
\begin{equation*}
\left.\Psi(z)\right|_{z=a}=1\quad\mathrm{and}\quad\left.\frac{\partial\Psi(z)}{\partial z}\right|_{z=a}=ik_2\,.
\end{equation*}
As the next step, we numerically integrate the Schr\"odinger equation from $z=a$ to $z=0$, while the wave function and its derivative need to satisfy
\begin{equation*}
\Psi(0)=A_1+A_\mathrm{R}\quad\mathrm{and}\quad ik_1A_1-ik_1A_\mathrm{R}=\left.\frac{\partial\Psi(z)}{\partial z}\right|_{z=0}\,.
\end{equation*}
such that
\begin{align*}
&A_1=\frac{\Psi(0)}{2}+\frac{i}{2k_1}\left.\frac{\partial\Psi(z)}{\partial z}\right|_{z=0}\quad\text{and}\\
&A_\mathrm{R}=\frac{\Psi(0)}{2}-\frac{i}{2k_1}\left.\frac{\partial\Psi(z)}{\partial z}\right|_{z=0}\,.
\end{align*}
To complete the one-dimensional problem, we numerically calculate the reflection and transmission coefficients $R$ and $\Theta_\mathrm{1D}$ using
\begin{equation*}
\Theta_\mathrm{1D}=1-R=1-\frac{A^\dagger_\mathrm{R} A_\mathrm{R}}{A^\dagger_1 A_1}\,.
\end{equation*}
We calculate the one-dimensional transmission because only momentum components perpendicular to the barrier contribute. Still, to determine the realistic transmission in three dimensions, we have to factor in the three-dimensional dispersion relation of the impinging electrons \cite{Hansen2004}. First, we calculate the three-dimensional partial transmission at a given energy $E$ by numerically integrating
\begin{equation}
\Theta_\text{3D}^\text P(E,V)=\frac{1}{E}\int_0^E\Theta_\text{1D}(E_\mathrm{z},V)dE_\mathrm{z}
\label{eq:t3d}
\end{equation}
with the applied voltage $V$ and the kinetic energy component $E_\mathrm{z}=E_1=\frac{\hbar^2k_1^2}{2m}$ of electrons corresponding to their initial motion in the $z$ direction. To determine the full three-dimensional transmission $\Theta_\text{3D}(V)$ we have to integrate $T_\text{3D}^\text P(E,V)$ over all relevant energies. We first calculate the transmission at zero temperature
\begin{equation*}
\Theta_0(V)=\frac{1}{e V E_\mathrm{F}}\int_{E_\mathrm{F}-e V}^{E_\mathrm{F}}\Theta_\mathrm{3D}^\text P(E,V)EdE
\end{equation*}
and then, using the Sommerfeld expansion valid for $\kb T\ll eV\ef$ \cite{Ash1976}, the lowest order correction term for a finite temperature
\begin{equation*}
\Delta \Theta(V)=\frac{\pi^2}{6}\frac{(k_\mathrm{B}T)^2}{e V E_\mathrm{F}}\,\left[\Theta_\mathrm{3D}^\text P(E_\mathrm{F},V)-\Theta_\mathrm{3D}^\text P(E_\mathrm{F}-e V,V)\right]
\end{equation*}
such that
\begin{equation*}
\Theta_\text{3D} (V)\simeq \Theta_0(V) + \Delta \Theta(V)\,.
\end{equation*}
After numerically computing
\begin{equation*}
\frac{d}{dV}\Theta_0V|_{V=0}\quad\mathrm{and}\quad\frac{d}{dV}\Delta \Theta V|_{V=0}
\end{equation*}
we calculate the linear-response two-terminal resistance with
\begin{multline}
R_\text{2TM}=2R_\mathrm{C}+R_\mathrm{NW}\\=2\left\{G_S\left(\frac{d}{dV}\Theta_0V|_{V=0}+\frac{d}{dV}\Delta \Theta V|_{V=0}\right)\right\}^{-1}+R_\mathrm{NW}
\label{eq:rtot}
\end{multline}
where
\begin{equation*}
G_S=\frac{\pi A}{\lambda_F^2}G_\text Q\quad\mathrm{with}\quad A=0.5\cdot 2\pi r L\,.
\end{equation*}
Here, $G_\text Q=2e^2/h$ is the conductance quantum, $r$ the radius of the NW and $L$ the contact width along the NW, while $\lambda_F$ is the Fermi wavelength (cf. Tab.~\ref{tab:values}).
\begin{table}[ht]
\begin{center}
\begin{tabular}{ccccc}
\hline
\hline
\textbf{parameter} & \textbf{value} &~~~& \textbf{parameter} & \textbf{value}\\
$m_e$ ($m_0$) & 0.067	&~~~~~~&	$\gamma$ (eV$^{-1}$) & 27.0794\\
$E_F$ (meV) & 14.0	&~~~~~~&	$\lambda_F$ (nm) & 24\\
$d$ (nm) & 13.0		&~~~~~~&	$r$ (nm) & 90\\
$\phi_0$ (eV) & 1.0	&~~~~~~&	$L$ ($\mu$m) & 1\\
$V$ (V) & 0.01 .. 1	&~~~~~~&	$R_\mathrm{NW}$ (k$\Omega$) & 25\\
$D_0$ (m$^2$/s) & 10$^{-4}$ &~~~~~~& $E_\mathrm{A}$ (eV) & 1.445\\
$c_\mathrm{Si}$ (m$^{-3}$) & 9$\times$10$^{24}$ &~~~~~~& $c_\mathrm{Ge}^\mathrm{b}$ (m$^{-3}$) & 4.42$\times$10$^{28}$\\
$z_0$ (nm) & 50 &~~~~~~& $\varepsilon_\mathrm{GaAs}$ & 13.18\\
\hline
\hline
\end{tabular}
\end{center}
\caption{Values employed for the calculations.\label{tab:values}}
\end{table}
We declared convergence to be achieved for $\delta V<0.1\,$mV. To simulate Schottky barriers in the present work, we assumed the nominal Schottky barrier width $W_\text{SB}(z_0)$ to be equal to the actual barrier width $a$. 

\subsection*{Analytical solution for a parabolic barrier}\label{app:parabolic}
\begin{figure}[ht]
\begin{center}
\includegraphics[width=.9\linewidth]{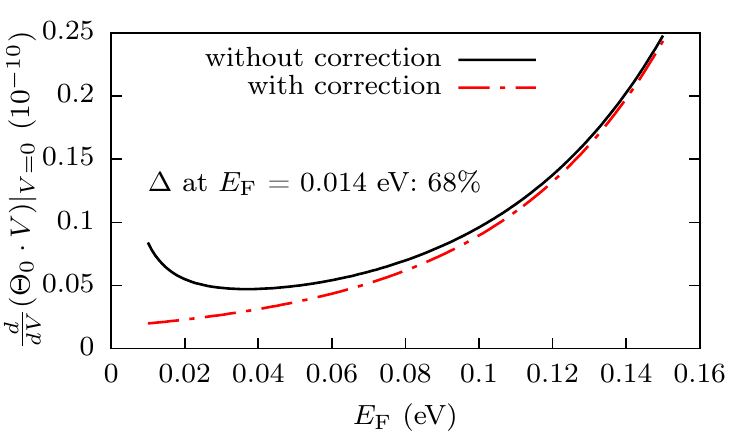}
\end{center}
\caption{The linear-response resistance at zero temperature of a barrier is proportional to $\frac{d}{dV}\Theta_0V|_{V=0}$, which is plotted for a parabolic barrier as a function of the Fermi Energy $E_F$. The red dashed line is the exact analytical solution. The solid black line neglects the two non-exponential terms in \eq{eq:ddvt0v}, as suggested as an approximation in Ref.\ \cite{Hansen2004}. \label{fig:ddV}}
\end{figure}

The transmission through a parabolic barrier of the form $\phi(z)=\phi_0\cdot(1-4(z/a-0.5)^2)$ can be calculated analytically, which we have used to verify the correctness of our numerical calculations. We follow the approach of Hansen and Brandbyge~\cite{Hansen2004}, but demonstrate that a term omitted in their calculation becomes relevant for small Fermi energies typical for doped semiconductors. We have to solve the integral in \eq{eq:t3d} to obtain the partial transmission $\Theta_\text{3D}^\text P$, where we take $\Theta_\text{1D}$ from Eq.\ (19) in Ref.\ \cite{Hansen2004}
\begin{equation}
\Theta_\text{1D}(E_z,V)=\frac{1}{1+\exp[\gamma(\tilde{\phi}_V-E_z)]}
\end{equation}
$$\mathrm{with}~~\gamma=\frac{\sqrt{2}\pi^2}{\mathrm{h}}\sqrt\frac{m_e}{\phi_0}d
~~\mathrm{and}~~\tilde{\phi}_V=\phi_0\left(1+\frac{1}{4}\frac{eV}{\phi_0}\right)^2\,.$$
Here, we define $eV$ as in Fig.\ 2 of Ref.~\cite{Hansen2004} to be \emph{positive}.
We obtain:
\begin{multline*}
\Theta_\text{3D}^\text P(E,V)=\frac{1}{E}\int_0^E\frac{1}{1+\exp[\gamma(\phi_V-E_z)]} dE_z =\\ 1 + \frac{1}{E\gamma}\left\{\ln[1+\exp(-\gamma(E-\phi_V))]\textcolor{red}{-\ln[1+\exp(\gamma\phi_V)]}\right\}\,,
\end{multline*}
where the term omitted in Eq.\ (21) of Ref.~\cite{Hansen2004} is marked red and
\begin{widetext}
\begin{multline*}
\Theta_0(V)=\frac{1}{e\cdot V  E_F}\int_{E_F-e\cdot V}^{E_F}E\Theta_\text{3D}^\text P(E,V)dE\\=
\frac{\mathrm{Li_2}[-\exp(-\gamma(\tilde{\phi}_V-E_F)]-\mathrm{Li_2}[-\exp(-\gamma(\tilde{\phi}_V-e\cdot V-E_F))]}
{\gamma^2 E_F e\cdot V}
\textcolor{red}{+\frac{\gamma\tilde{\phi}_V-\ln[1+\exp(\gamma\tilde{\phi}_V)]}{\gamma E_F}}\,.
\end{multline*}
\end{widetext}
To obtain the resistance at zero temperature, we compute
\begin{multline}
\frac{d}{dV}\Theta_0V|_{V=0}=\frac{1}{\gamma E_F}\left\{\exp[-\gamma(\tilde{\phi}_V-E_F)]\right.\\
\left.\textcolor{red}{+\gamma\tilde{\phi}_V-\ln(1+\exp[\gamma\tilde{\phi}_V])}\right\}\label{eq:ddvt0v}\,.
\end{multline}
In \fig{fig:ddV}{} we evaluate the impact of the neglected term in Ref.~\cite{Hansen2004} (second line, red) by plotting $\frac{d}{dV}\Theta_0V|_{V=0}$ as a function of the Fermi energy $E_F$. While omitting the red terms yields a good approximation above a Fermi energy of about 100\,meV, the correct value of $\frac{d}{dV}\Theta_0V|_{V=0}$ is almost 70\% smaller at the Fermi energy of 14\,meV in our nanowires. All other parameters required for this comparison are listed in Tab.~\ref{tab:values}. Omitting the correction term, as proposed in Ref.~\cite{Hansen2004} clearly yields an incorrect description of our system.
With
\begin{equation}
\frac{d}{dV}\Delta \Theta V|_{V=0}=\frac{\pi^2}{6}\frac{(k_\mathrm{B}T)^2}{E_F}\frac{\gamma\exp[\gamma(\tilde{\phi}_V-E_F)]}{(1+\exp[\gamma(\tilde{\phi}_V-E_F)])^2}
\label{eq:ddvdtv}
\end{equation}
and using Eqs.\ (\ref{eq:ddvt0v}) and (\ref{eq:ddvdtv}), the 2TM resistance can be calculated from \eq{eq:rtot}.

\begin{figure}[ht]
\begin{center}
\includegraphics[width=1\linewidth]{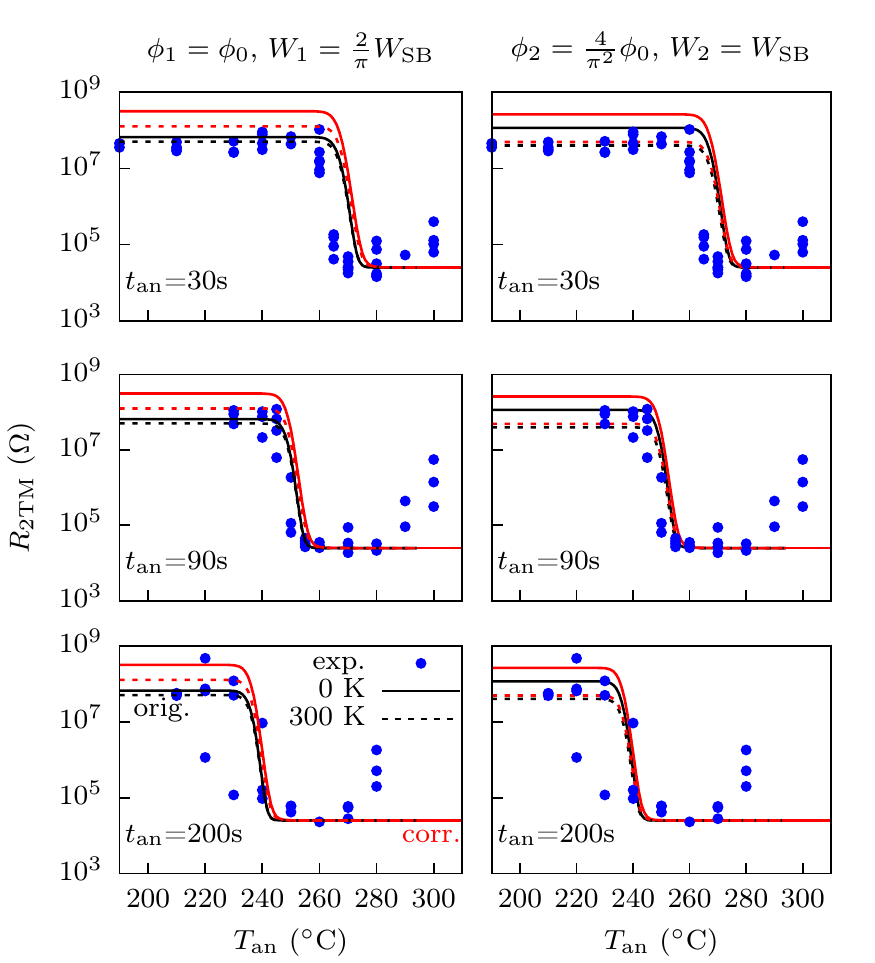}
\end{center}
\caption{Two terminal resistance $R_\mathrm{2TM}$ calculated for temperatures $T=0$\,K (solid) and 300\,K (dashed) for different annealing times as a function of the annealing temperature. Analytical solutions for parabolic barriers: left hand side for fixed barrier height \{1\}, right hand side for fixed barrier width \{2\}. Black curves are obtained without the correction term in \eq{eq:ddvt0v}, red curves include the correction. $R_\mathrm{2TM}$ values measured at room temperature are shown as blue dots.\label{fig:rtemp}}
\end{figure}

In the following, we approximate the Schottky barrier with barriers of parabolic shape in order to use the formalism introduced by Hansen and Brandbyge. For a comparison, we keep either \{1\} the barrier height $\phi_0$ or \{2\} the barrier width $W_\mathrm{SB}$ unchanged and adjust the other parameter to keep the tunnel probability at fixed Fermi energy. The width depends on the Ge concentration according to \eq{eq:schott} which, in turn, is a function of the annealing temperature $T_\mathrm{an}$ as shown in \fig{fig:aVsT}{}. The barriers are approximated as
\begin{equation}
[E-E_F]_1(z)=\phi_\mathrm{0}-\frac{\phi_\mathrm{0}\pi^2}{W^2_\mathrm{SB}}\left(z-\frac{W_\mathrm{SB}}{\pi}\right)^2
\label{eq:par1}
\end{equation}
for \{1\} and as
\begin{equation}
[E-E_F]_2(z)=\frac{4\phi_\mathrm{0}}{\pi^2}-\frac{16\phi_\mathrm{0}}{(W_\mathrm{SB}\pi)^2}\left(z-\frac{W_\mathrm{SB}}{2}\right)^2
\label{eq:par2}
\end{equation}
for \{2\}.
The respective heights and widths of the two parabolic barriers are
\begin{equation}
\phi_1=\phi_\mathrm{0}~~\mathrm{and}~~W_1=\frac{2W_\mathrm{SB}}{\pi}
\label{eq:hw1}
\end{equation}
\begin{equation}
\phi_2=\frac{4\phi_\mathrm{0}}{\pi^2}~~\mathrm{and}~~W_2=W_\mathrm{SB}\,.
\label{eq:hw2}
\end{equation}
Replacing $\gamma$ and $\tilde{\phi}$ in \eq{eq:ddvt0v} with
\begin{equation}
\gamma_{1,2}=\frac{\sqrt{2}\pi^2}{h}\sqrt{\frac{m_e}{\phi_{1,2}}}W_{1,2}
\end{equation}
and $\phi_{1,2}$, respectively, we compute the 2TM linear-response resistance R$_\text{2TM}$ of the nanowire and two contacts using \eq{eq:rtot}. 
In \fig{fig:rtemp}{} we present the calculated $R_\text{2TM}$ as a function of the annealing temperature for different annealing times for the two parabolic barriers with and without the correction terms in \eq{eq:ddvt0v}. For a direct comparison we also plot the measured $R_\text{2TM}$.

\FloatBarrier

\nocite{apsrev41Control}
\bibliographystyle{apsrev4-1_mod}
\bibliography{literature}

\end{document}